# Modelling and analysing relations between entities using the multi-agent and social network approaches

Jarosław Koźlak, Anna Zygmunt, Edward Nawarecki
Department of Computer Science
AGH University of Science and Technology
Al. Mickiewicza 30, 30-059 Kraków, Poland
e-mail: {kozlak, azygmunt, nawar}@agh.edu.pl

***Abstract*— In this work, the concept of a system for analysing social relations between entities using the social network analysis and multi-agent system approaches is presented. The following problems especially appear within the domain of our interests: identification of the most influential individuals in a given society, identification of roles played by the given individuals in that society and the recognition of groups of individuals strongly connected with one another. For the analysis of these problems, two application domains are selected: an analysis of data regarding phone calls and analysis of Internet Weblogs.**

## I. INTRODUCTION

The goal of our work is to distinguish groups functioning in society and to identify the reasons that cause their formation and what motivates entities to join them. In particularly, we are interested in analyzing the dynamics of these groups over a defined period of time. To analyse such groups, several problems have to be solved. The first is to determine the borders of the given group and decide if a given entity belongs to it in a given time or not. The second is to decide the nature of interpretation of the interactions between subjects and how long the social relations lasts. To analyse the group organisation, it is also useful to have an estimation of the importance of any given individual subject, either in the context of a given society or in the context of a given role and how roles played by the individual should be interpreted.

We assume the representation of the problem domain by a graph where the subjects/individuals are the nodes and the edges represent the social relations which are the effects of different kinds of interactions.

In order to analyse such a problem, we decided to use methods and solutions offered by two different domains: the social network analysis (SNA) and the multi-agent approach. The SNA approach provides measures which make it possible to determine the importance of different nodes in the network, as well as to distinguish sub-societies/groups in it and to analyse their characteristics (such as, stability in time, closeness/density, etc).

The analysys of social networks and the problem of group organisation in particular, has an important impact with respect to counteracting crime. It makes it possible to identify the most important elements of criminal organisations or main propagators of dangerous or criminal ideas. In this, paper we presented two application domains: an analysys of data concerning phone calls and analysis of Internet weblogs. The application of the approach to phone calls analysis makes it possible to identify the roles in a criminal organisation and identifies the most dangerous group members. The analysis of the Internet weblogs may have various application. Firstly, it allows us to analyse the groups which propagate felonious (criminal) ideas such as fascism, peadophilia or terrorism on blogs and facilitaties the identification of the most important members of these groups. Secondly, this methodology my be also applied in the analysys of less structured data such as the whole Web - pages and relations between them.

In the second section, a general description of the research domain is presented. It contains information about methods of network dynamic analysis, identification of groups and different methods of describing the importance of individuals in the observed society. Section three contains the description of the model of the used approach. The subsequent two sections contain a descriptions of the realised systems and scenarios of their applications, together with selected results. Section six concludes.

## II. DOMAIN OVERVIEW

The problem domain embraces problems of identifying roles of social entities (especially the most important and influential), analysis of social network dynamics as well as recognition of groups/societies and how they evolved.

### *A. Social Network Dynamics*

The analysis of the social network dynamic is a subject of a lot of research being carried out today. Different measures of describing changes of the analysed society, which was a criminal organisation, are proposed in [18]. These measures are: centrality of nodes, density, cohesion and stability of groups. The analysis described in [13, 18] concerned the illegal trafficking of drugs and a terrorist network (Al Qaeda).

The approaches for analysing social network dynamics are classified in [18]. The authors distinguish between descriptive methods, statistical methods and simulation methods. The objective of the descriptive methods is to find changes of the network structure, analysis of the conformity of the analysed empirical data and verified sociological assumptions. The statistic methods describe changes of the network and explain the reasons of these changes, the changes come from the stochastic processes, such as reciprocity, transitivity and balance, determining the network behaviour. In [17] a wider overview of the statistic methods is given. In particular, the models of the network can be considered as continuous-time Markov chain models. The simulation methods are based on the application of the Multi-agent approaches to the analysis of the network dynamics.

The work [6] uses the Hidden Markov Models for identifying subgroups with suspicious behaviour in a social network. The goal is to analyse criminal activities.

An idea of a multi-agent system used for the analysis of social networks is presented in [11]. The approach is based on the assumption that each user in the network tries to optimise its individual utility. This utility is dependent on the existence of the connections with other users with a given configuration. The existence of each connection on the one hand increases the utility of the user but on the other, the user has to incur costs which lead to a decrease in utility. Such networks evolve up to the moment of achieving a maximum global utility.

Various systems for analysing static and dynamic social networks with the use of different approaches have been developed. The interesting attempt of constructing an environment which integrates features of different systems of these kinds, either cooperating with them and using their analysis or exchanging data with different formats with them is described in [4].

In our analysis, in addition to values of measures, the information about the roles attributed to the neighbouring nodes was taken into consideration. Alongside this, tests were carried out to detect the time of occurrence of important events affecting the organization, and for this we used the control cards method (cards preservation process), and in particular CUSUM card counting based on the cumulative sum of parameters describing the process [10]. The concept of applying the CUSUM approach for analysing the dynamics of social networks was proposed in [13].

### *B. Group Dynamic*

Important elements in analysis are identifying substructures and cliques as well as how these structures evolved in time. In [18], apart from the classical measures which describe nodes in SNA, the group level measures, such as link density, group cohesion and cluster stability are introduced. The link density describes the completeness of connections between group members and is expressed as a quotient of the sum of connections existing between the members of the group to the maximum possible number of connections between them (which means that each member of the group has connections with every other member of the group). Group cohesion informs us if the relations of members of the group one with another are stronger than their relations with nodes outside the group. The measure is calculated as a quotient of the average number of connections of the group members with other group members to the average number of its connections with the nodes not belonging to the group. Finally, the third measure, group stability, describes how the group is calculated in time. The stability of the group from the time point $t_1$ to the time point $t_2$ is defined as a quotient of the quantity of the set being the common part of the two sets which contain elements of this group in times $t_1$ and $t_2$ to the quantity of set being the sum of these sets.

### *C. Influence and Trust*

An important element describing dependencies among entities (members of groups) is the analysis of influences exerted by given nodes in the network on one another. To achieve this goal, different measures typical for social network analysis or estimations used in the multi-agent approach might be applied.

Some measures used in the social network approach directly offer a possibility to state the importance of given nodes and to find a way of how they exert the influence on others. The very simple measure of importance of the node is associated with its degrees – numbers of incoming and outgoing links. The incoming degree (described by the number of links incoming to the degree) describes the attraction that arises, and may represent influence or reputation.

The outgoing degree (described by the number outgoing links) represents the level of activity showed by the node. Another important measure which shapes interactions is trust, which is analysed for example in [15].

### *D. Reputation*

The problems of defining reputation and of finding measures which express their features is widely explored in the multi-agent systems domain. In [8] an expansive overwiew of reputation definitions and methods of estimation both in the society and multi-agent systems is given. In this work, currently accepted reputation definitions are also presented. In one of the approaches, the reputation is identified with the history, for example knowledge about previous behaviour of the interaction partners. In other approaches, reputation means that the individual is positively perceived. In economic theories, the reputation may be considered as a factor associated with advertisements, which has influence on the level of sale and profits, while in definitions arising from game theory, the reputation is information about the behavior of interaction partners coming from third parties. Reputation may also be considered as an evaluation of future profits/advantages which will be the results of cooperation with any given entity.

## III. MODEL OF RELATIONS BETWEEN ENTITIES

For this analysis we elaborated the model of the multi-agent system, which takes groups and dependencies between them into consideration. Below, we will define three of the most fundamental concepts of our model: agent society, basic agent representing given entities (in the analysed case – given bloggers) and the agent representing the group of strongly connected individuals.

The society of agents may be described as follows:

S = (AS, I, Org, M, P, R, RI),

where:

AS – is a set of all agents,

I – a set of interactions between agents: I: $AS \times AS \rightarrow N$,

Org – organisation which is a criterion to distinguish groups from the society,

M – the matrix which contains values of social network measures for all the agents ,

P – set of schemes describing roles which exist in the system,

R – set of roles,

RI – function which assign roles in the Social Network, played by the given agent.

RI: AS → R

The description of organisation (Org) may be established using different methods. The selected method, based directly on the interaction scheme, is the description of the quantity of connections which took place between the pairs of agents in the analysed period of time. The matrix M is made up of $M_i$ rows, where the rows describe vectors of measures for the given node. The elements of this vector, $m_{ij}$, contains values of different measures for the given node, such as degrees of incoming and outgoing links, centrality betweeness, bary centre centrality, HITS, Page Rank, Markov Centrality etc.

An individual $A_i$ agent is represented by entity

$$A_i = (N_i, EN, M_i, R_i, T_i) \quad A_i \in AS$$

where:

Ni – set of neighbours (agents with which Ai has direct interactions), $N_i \subset AS$

EN – the functions evaluating the strength of the connection with each agent from the neighbourhood $N \rightarrow (0,1)$

$M_i$ – vector of measures for the given node,

$R_i$ – the role assigned to a given node,

$T_i$ – the parameter describing a character of given agent activities. For example, for the phone calls domain it may describe a profile of behaviour of a given user: from which location the user usually calls, in which time periods, frequent sequences of calls in which this user participates.

For the blogs domain, it may include subjects of posts and comments of the agent, described by a set of tags of the posts or by a result of the analysis of the post contents (for example, expressed by statistic measures using the vector model which describe the frequencies of the appearance of words/subjects in the posts),

The Agent-group is represented by a tuple

$$AG_j = (f, e, s, S),$$

where:

f –function which represents strength of membership of the agent in the given group

f: $a \in AS \rightarrow (0,1)$

e – the function which classifies agents on the basis of the strength of the membership to the group. As results, the agent may obtain four different degrees: membership in the centre of the group – core, membership in the group weaker related to the group (circumjacent), weak/accidental relations with the group or lack of this relation ,

e: f(s) -> (kernel, circumjacent, weak, not related)

s: strategy which describe reasons of the group formation and goals of their activities, for example, described by the values of the measures in the group or kinds of themes/subjects discussions.

S – set of possible strategies

**Model Input and Output:**

*Input:* Criteria of indentifying groups. Different algorithms for extracting societies may be used here, for example, described in [1], as well as different configurations of these algorithms and different assumed thresholds.

*Output:* Identification of the most influential individual and groups with their characteristics and description of their evolution, identified reasons for joining groups by individuals (increase of measures, desired relations with subject profiles), dependencies between groups, group inclusions and intersections.

**Analysis of evolution** of the state of the society may be carried out in different aspects:

- Change of the state of the society - collective comparison of the time dependences of the measures of the given basic agents.
- Change of the state of the agent - the comparison of the time changes of the measures of the given agent, comparison of the state of the neighbourhood set and strengths of relations with the neighbours as well as the change of roles of the agents.

- Change of the state of the group – change of the values of the membership functions to the given group, change of sets describing the kernel, circumjacent and weekly related agents, change of the measures describing the stability of the group as well as change of the group strategy.

## IV. SHORT DESCRIPTION OF THE COMPUTATION ENVIRONMENTS

In the framework of the work carried out, environments for the analysis of data which may be represented using the social network were developed. In the analysis the classic measures of social network such as Bary Center, Betweenness Centrality, Degree In Centrality, Degree Out Centrality, Hubness, Authoritativeness, Page Rank, Markov Centrality were used as well as the algorithms for the identification of groups in the society.

### *A. Analyis of phone calls*

The schema of system functionality for the analysis of phone calls is presented in fig. 1 [12,14,19]. The input data (information about phone calls and base stations), methods of configuration algorithm (by the definition and description of the set of roles) as well as the choice of the method of the visualization of results. The system provides the results in the form of a graph which represent the social network or its part as well as the measures of the nodes individuals and roles assigned with them. Analysis for the selected time periods are also possible.

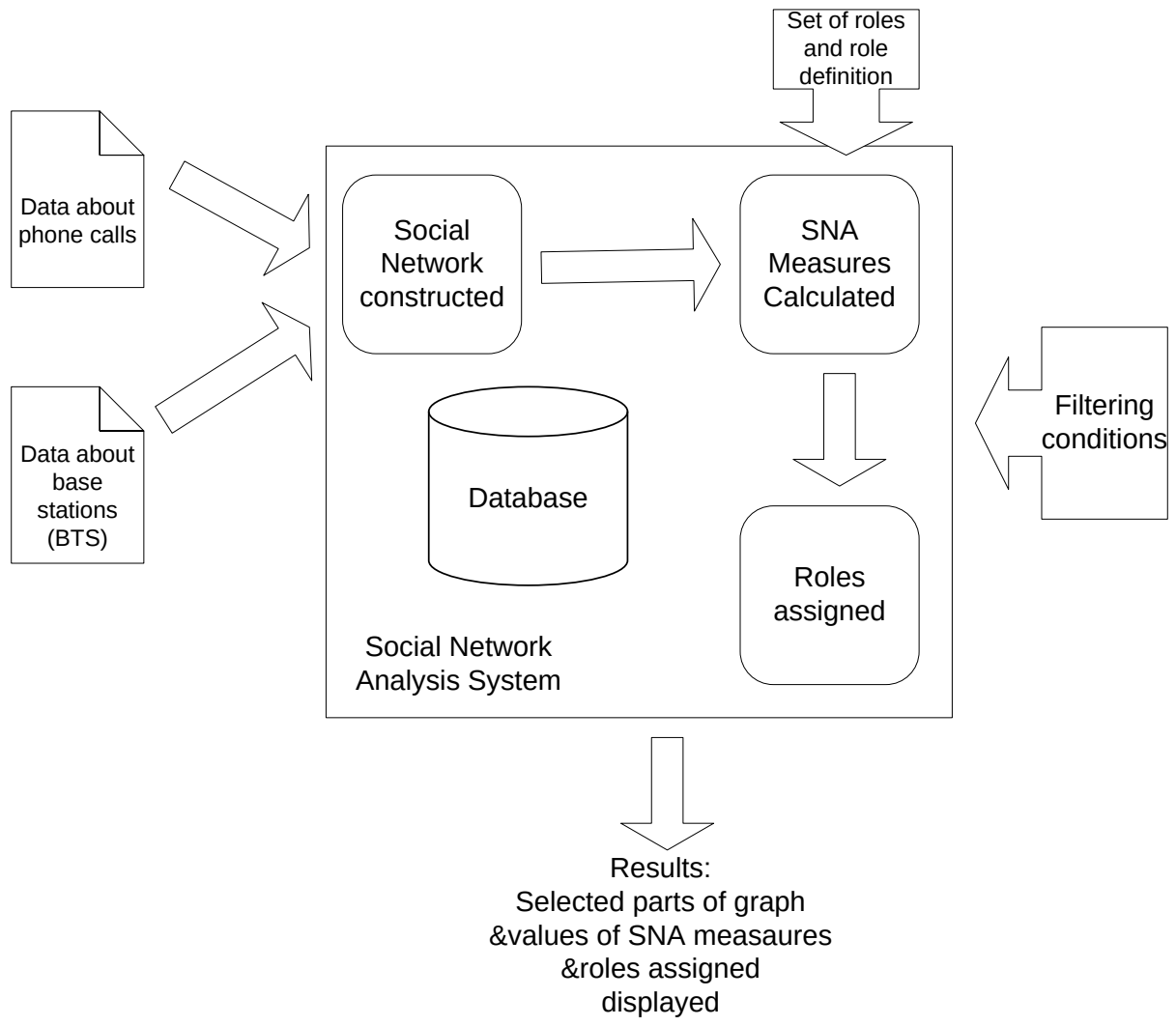

Fig. 1. System analysing data about phone calls using SNA

### *B. Analysis of blogs*

In our work related to Weblogs, we focused on the analysis of blogs from the www.salon24.pl portal [20]. The environment (fig. 2.) for performing these analyses was developed. This environment contains a module, which collects the contents of blogs and stores it in databases while crawling the web. The performed analysis embraced the computation of SNA measures for static (on the basis of the full set of data) or dynamic (the analysis of the state of the network in subsequent time periods). The text content of the posts was also analysed using the statistical representation of words/concepts in the texts and times when posts or comments were written and applying the method called term frequency – inverse document frequency (TF-IDF) measure. To achieve it, documents (posts or comments) are described by the vector of weights associated with the words appearing in the analysis. The weights were calculated using a TF-IDF method, their values depending on the number of times any given word appears in the document and the degree of how much the word distinguishes a document from other analysed documents in the considered set (the lower the number of appearances of a word in a document, the stronger it distuingushes the given document from other documents). To find documents similar to any given post/set of comments, we used a cosine distance between the vectors representing the documents in question and other documents.

The last domain of analysis was the identification of groups and analysis of their evolution.

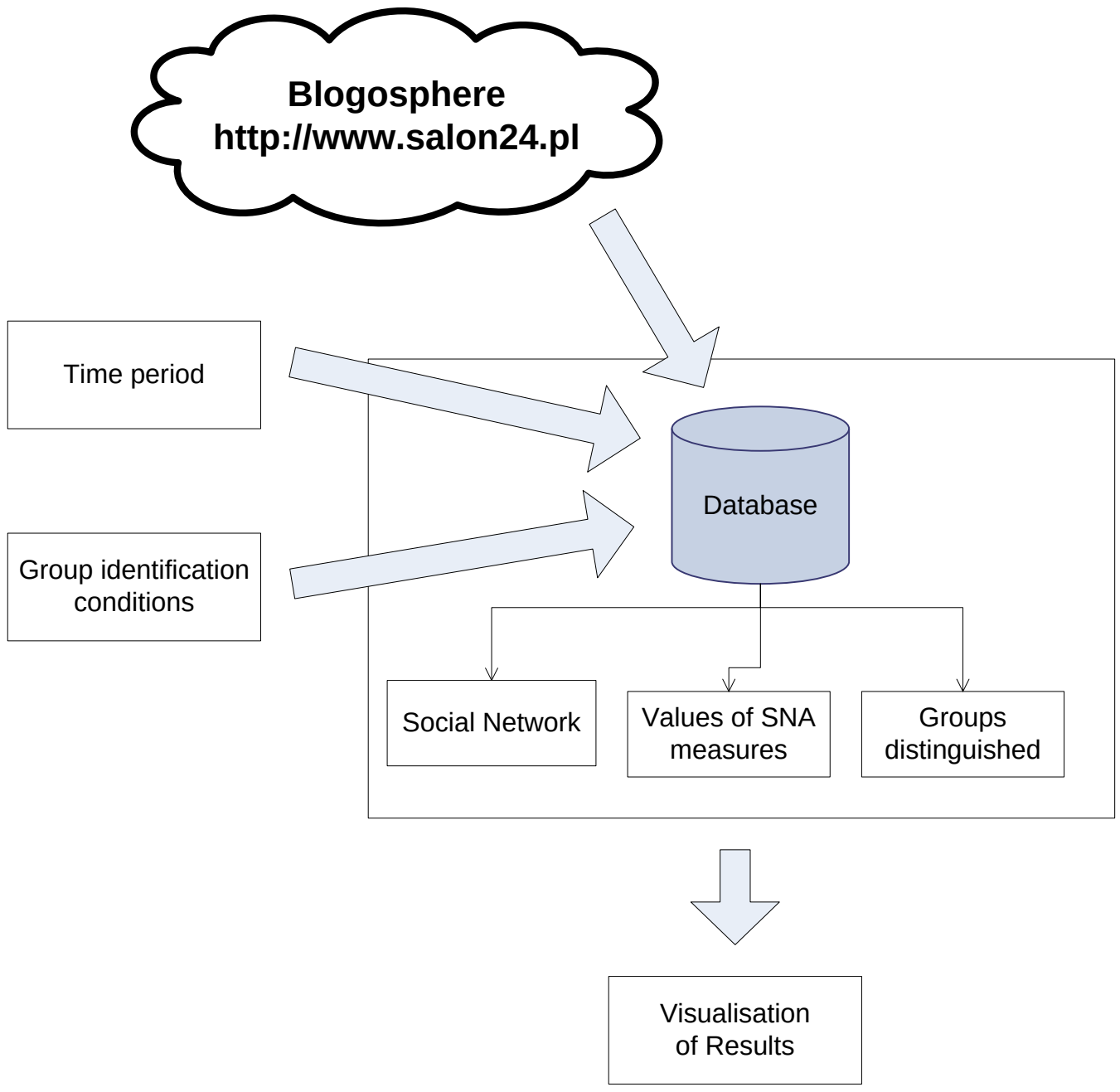

Fig.2. System analysing data about phone calls using SNA and MAS approach

## V. ANALYSIS OF THE EXAMPLES OF SCENARIOS

Below, two selected scenarios of application of the developed systems are presented. The first one concerns the analysis of data about phone calls which assigns the individuals with roles played by them and the second one concerns the identification of the most influential bloggers.

**Analysis of phone calls.** In its current stage, our approach is based on the calculation for each node in the network, which represents phone users, values of various measures, either classical measures used in the Social Network Analysis, or special measures, dependent on the problem domain, introduced by us. The second group of measures, the domain dependent measures, are: mobility, spatial range of incoming and outgoing calls, length of calls, average number of incoming/outgoing calls for one day, average number of incoming/outgoing SMSs, number of different incoming interlocutors, number of different outgoing interlocutors, calls/SMSs ratio, time period of activity in the network

As a demonstrative illustration of the functioning of the system, a case with 133197 records of data about connections between 7757 interlocutors was selected. The full information about the calls was provided for 24 phone numbers under surveillance. The investigation concerned the activities of the criminal group which performed a car theft. On the basis of this data a social network was generated. The subsequent analysis results in being able to determine the values of the given parameters, which allows us the possibility to obtain a preliminary identification of roles played by any given person.

Table 1 contains the description of roles definitions on the basis of the selected SNA measures. Accordingly, 0-2 means that the role needs a very low value of the given measure, 2-4 – low value, 4-6 average value, 6-8 high value and 8-10 – very high value. The nodes with the roles of Organiser, Receiver and Soldier were interesting for us as the first two of these roles indicated the most important organization members. In fig. 3 a selected part of the social network with the most interesting nodes and connections between them is presented. The system identified 5 nodes with the roles of Organiser and 6 nodes with the role of Receiver, the real Organisers and Receivers were among them.

Table 1: Description of role definitions

| Role<br>Meeasure | Organiser | Receiver | Soldier | Outsider |
|---|---|---|---|---|
| Bary Center | 2-4 | 2-4 | 2-4 | 0-2 |
| DegreeIn | 4-6 | 8-10 | 2-6 | 0-2 |
| DegreeOut | 0-2 | 2-4 | 2-4 | 0-2 |
| Hubness | 0-2 | 0-2 | 0-2 | 0-2 |
| Authoritativeness | 8-10 | 0-2 | 0-2 | 0-2 |
| Page Rank | 2-4 | 6-8 | 0-2 | 0-2 |
| Betweeness Centrality | 4-6 | 2-4 | 0-2 | 0-2 |
| Markov Centrality | 6-8 | 6-8 | 0-2 | 0-2 |

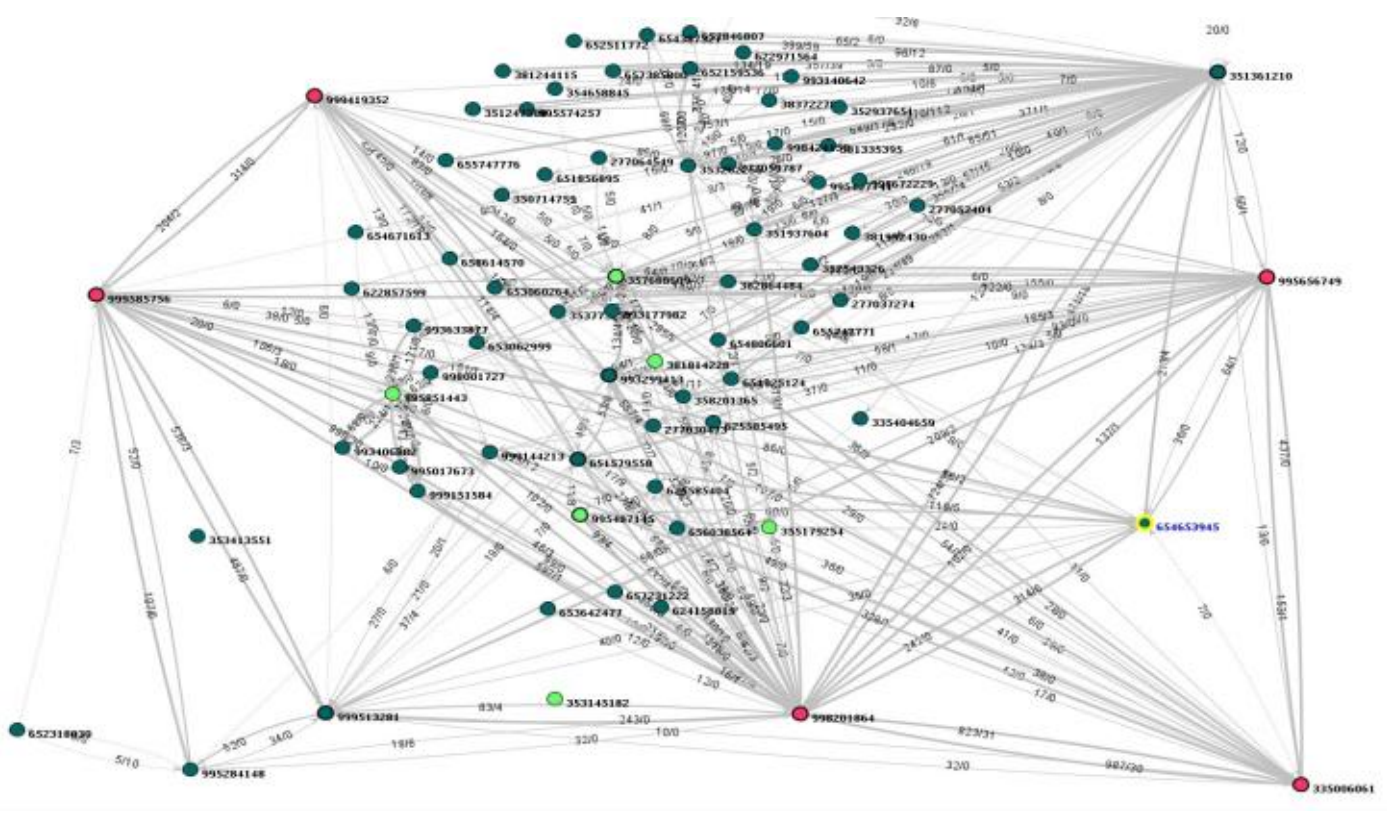

Fig. 3. Graph of connections obtained after the processing of data about phone calls.

**Internet blogs.** As a second application domain for our analysis concerning the modelling of relations between entities, dynamics and group characteristics, we selected the Internet blogosphere. It is an interesting domain for several reasons such as: open access to a wide range and rich source of data, access to not only information about inter-connections, but also to the data exchange between entities as well as the high quantity and dynamics of interactions. Additionally, the obtained analysis can be possibly interpreted. After observing blogs, it is relatively easy to establish which bloggers have an important impact on group opinions, which bloggers have similar interests and opinions on a given subject and which opinions are controversial.

The different analysis of the node characteristics and the methods of group formation for the blogs from http://www.salon24.pl were carried out. Below, the selected experimental scenarios will be presented. The goal was to identify the individuals with the highest influence on other bloggers. To achieve this goal, the following simple measurements may be taken into consideration:

- degree of incoming – informs how many blogs point out to the given blog, this measurement describes the position of the given blogger in the society (fig. 4.). Incoming links are calculated from the number of comments to the posts that are written on any given blog by other users.
- degree of outgoing– informs on how many blogs the given blog is linked to, it expresses the activity of the blogger in the society (fig. 5). The outgoing links are calculated from the comments written to the posts of other bloggers by any given blogger.

For a better visualisation of the characters of the nodes on the one figure, a new measurement was introduced: .

$$m_3 = \frac{m_1^2}{m_2}$$

where:

$m_1$ – degree of incoming of the node,

$m_2$ – degree of outgoing of the node.

The application of this measurement is to point out the nodes which have substantially more of incoming connections over outgoing ones and at the same time have a lot of incoming links. The performed estimation showed that there are some nodes which have very high values of this measurement (tens of thousands) and the most active and influential bloggers belong to this group (fig. 6.).

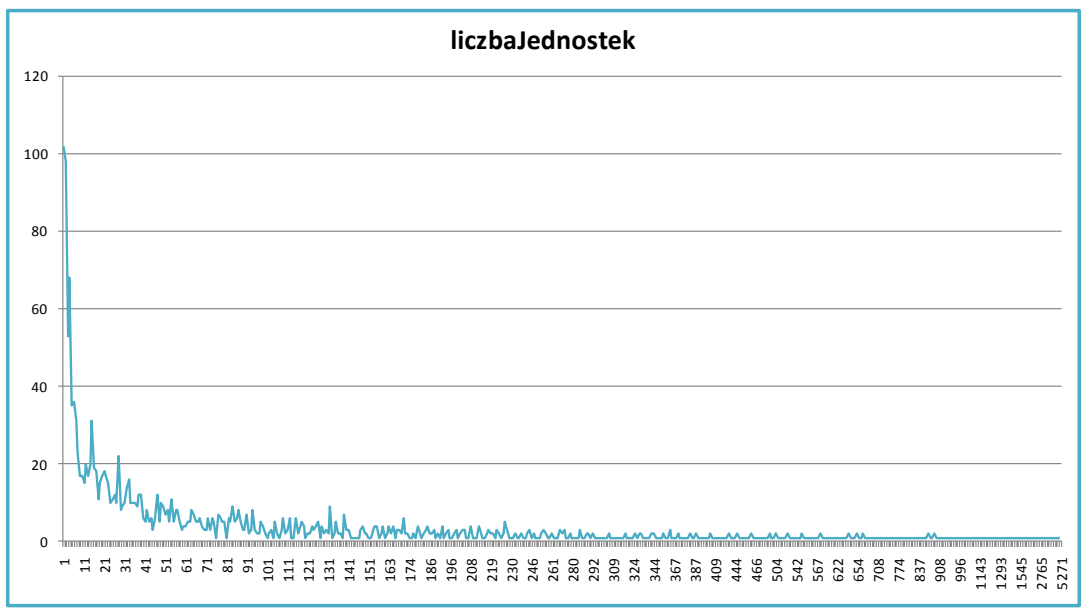

Fig. 4. Activity of nodes expressed by the number of outgoing connections. Numbers of nodes (y) with given values of outgoing connections (x).

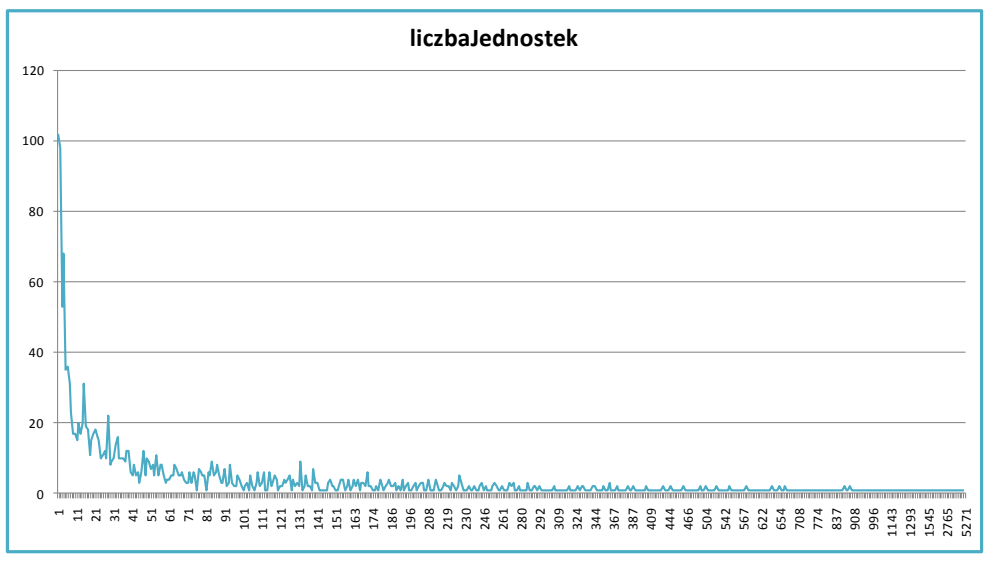

Fig 5. Authority expressed by the number of incoming connections. Numbers of nodes (y) with given values of outgoing connections (x).

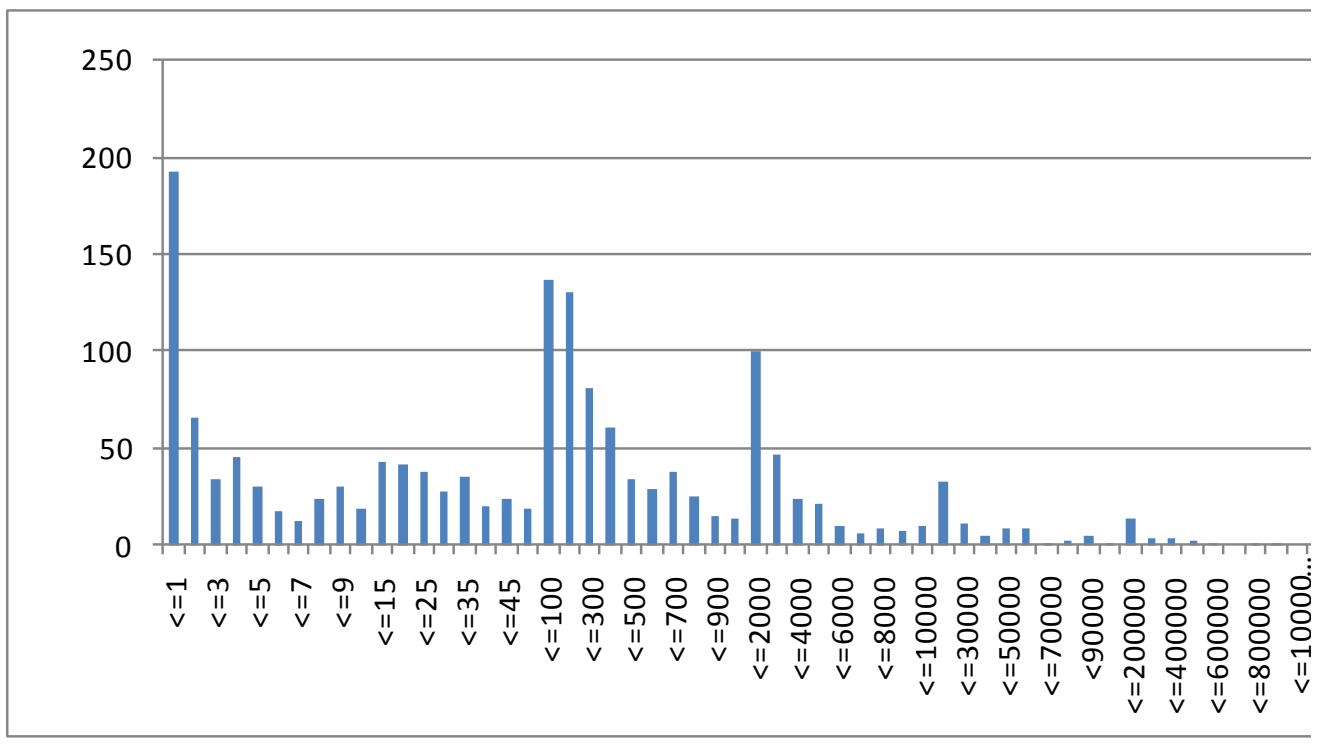

Fig. 6. Identification of nodes with a high authority with the use of measure $m_3$. Numbers of nodes (y) with value of the measure belonging to the given ranges(x)

## VI. CONCLUSIONS

In this paper, the concept and defined terminology used for the development of the computation environment for the analysis of social relations, roles and importance of individuals in the society was presented. We presented the environments for social analysis which especially include the possibility of performing such activities. We also described examples of scenarios of use of the environment and selected obtained results. Future work will focus on the analysis of group characteristics and dependences between them, considering the dynamic evolution of the organization.

**Acknowledgements.**

The research leading to these results has received funding from the European Community's Seventh Framework Program (FP7/2007-2013) under grant agreement n° 218086. The authors wish thank former and present M.Sc. students in the Department of the Computer Science AGH-UST (Łukasz Preiss, Tomasz Skucha, Lukasz Krupczak) for their participation in the development of the presented applications.